\documentclass[aps,prd,preprint,12pt,superscriptaddress,nofootinbib,a4paper]{revtex4-1}
\pdfoutput=1

\usepackage[utf8]{inputenc}
\usepackage{amsmath,amssymb}
\usepackage[separate-uncertainty=true]{siunitx}
\usepackage{graphicx}
\usepackage[usenames,dvipsnames]{xcolor}
\usepackage[caption=false]{subfig}
\usepackage{hyperref}
\usepackage[capitalize]{cleveref}
\usepackage{booktabs}
\usepackage{tabularx}
\usepackage{xspace}
\usepackage{color}
\usepackage[normalem]{ulem}
\usepackage{slashed}
\usepackage{bbold}
\usepackage{wasysym}
\usepackage{graphicx}
\usepackage{mathrsfs} 

\allowdisplaybreaks

\graphicspath{{graphics/}}

\newcommand{\eqn}{equation}
\newcommand{\lb}{\left(}
\newcommand{\rb}{\right)}
\newcommand{\be}{\beta}
\newcommand{\al}{\alpha}

\newcommand{\HSv}[1]{\texttt{HiggsSignals-#1}}
\newcommand{\HBv}[1]{\texttt{HiggsBounds-#1}}
\newcommand{\HS}{\texttt{HiggsSignals}}
\newcommand{\HB}{\texttt{HiggsBounds}}

\newcommand{\GeV}{{\ensuremath\rm GeV}}
\newcommand{\TeV}{{\ensuremath\rm TeV}}
\newcommand{\lam}{\lambda}

\newcommand{\pb}{{\ensuremath\rm pb}}
\newcommand{\fb}{{\ensuremath\rm fb}}
\newcommand{\ab}{{\ensuremath\rm ab}}

\DeclareSIUnit{\pb}{pb}
\DeclareSIUnit{\fb}{fb}
\AtBeginDocument{
\heavyrulewidth=.08em
\lightrulewidth=.05em
\cmidrulewidth=.03em
\belowrulesep=.65ex
\belowbottomsep=0pt
\aboverulesep=.4ex
\abovetopsep=0pt
\cmidrulesep=\doublerulesep
\cmidrulekern=.5em
\defaultaddspace=.5em

\newcolumntype{C}{>{\centering\arraybackslash}X}
\newcolumntype{b}{C}
\newcolumntype{s}{>{\hsize=.6\hsize}C}
\newcolumntype{R}{>{\raggedleft\arraybackslash}X}
}

\begin{document}
\bibliographystyle{hunsrt}
\date{\today}
\rightline{RBI-ThPhys-2022-9, CERN-TH-2022-041}
\title{{\Large New Physics with missing energy at future lepton colliders - Snowmass White Paper}}

\author{Jan Kalinowski}
\email{Jan.Kalinowski@fuw.edu.pl}
\affiliation{Faculty of Physics, University of Warsaw, ul.~Pasteura 5, 02--093 Warsaw, Poland}

\author{Tania Robens}
\email{trobens@irb.hr}
\affiliation{Ruder Boskovic Institute, Bijenicka cesta 54, 10000 Zagreb, Croatia}
\affiliation{Theoretical Physics Department, CERN, 1211 Geneva 23, Switzerland}

\author{Aleksander Filip \.Zarnecki}
\email{Filip.Zarnecki@fuw.edu.pl}
\affiliation{Faculty of Physics, University of Warsaw,  ul.~Pasteura 5, 02--093 Warsaw, Poland}

\renewcommand{\abstractname}{\texorpdfstring{\vspace{0.5cm}}{} Abstract}

\begin{abstract}
    \vspace{0.5cm}
  Two models that extend the particle content of the SM and provide dark matter candidates, namely the Inert Doublet Model and the Two-Higgs Doublet model with additional pseudoscalar,   are  confronted with current experimental and theoretical constraints and  predictions for production cross sections for various standard pair-production modes within these models at future lepton colliders are presented.
\end{abstract}

\maketitle


\section{Introduction}
We discuss two new physics models that extend the Standard Model (SM) particle sector by additional scalars and  provide a dark matter candidate, namely the Inert Doublet Model (IDM) and the Two-Higgs Doublet Model with additional pseudoscalar (THDMa). Both models are confronted with current theoretical and experimental constraints. From the theoretical side, these include the minimization of the vacuum as well as the requirement of vacuum stability and positivity. We also require perturbative unitarity to hold, and perturbativity of the couplings at the electroweak scale.

Experimental bounds include the agreement with current measurements of the properties of the 125 \GeV~ resonance discovered by the LHC experiments, as well as agreement with the null-results from searches for additional particles at current or past colliders. We also confront the models with bounds from electroweak precision observables (via $S,\,T,\,U$  parameters), 
B-physics observables $\lb B\,\rightarrow\,X_s\,\gamma,\,B_s\,\rightarrow\,\mu^+\,\mu^-,\,\Delta M_s\rb$, as well as agreement with astrophysical observables (relic density and direct detection bounds). We use a combination of private and public tools in these analyses, where the latter include HiggsBounds \cite{Bechtle:2020pkv},  HiggsSignals \cite{Bechtle:2020uwn}, 2HDMC \cite{Eriksson:2009ws}, SPheno \cite{Porod:2011nf}, Sarah \cite{Staub:2013tta}, micrOMEGAs \cite{Belanger:2018ccd,Belanger:2020gnr}, and MadDM \cite{Ambrogi:2018jqj}. Experimental numbers are taken from \cite{Baak:2014ora,Haller:2018nnx} for electroweak precision observables, \cite{combi} for $B_s\,\rightarrow\,\mu^+\,\mu^-$, \cite{Amhis:2019ckw} for $\Delta M_s$ and \cite{Planck:2018vyg} and  \cite{Aprile:2018dbl} for relic density and direct detection, respectively.
Bounds from $B\,\rightarrow\,X_s\gamma$ are implemented using a fit function from \cite{Misiak:2020vlo,mm}. Predictions for production cross sections shown here have been obtained using  Madgraph5 \cite{Alwall:2011uj}.

\section{The Inert Doublet Model}\label{sec:idm}
\subsection{The model}
The Inert Doublet Model (IDM) \cite{Deshpande:1977rw,Cao:2007rm,Barbieri:2006dq} is an intriguing new physics model. In this model, the SM scalar sector is enhanced by an additional $SU(2)\,\times\,U(1)$ gauge doublet $\phi_D$. As in the SM, the first doublet $\phi_S$  contains the  SM-like Higgs boson $h$, while the inert one  $\phi_D$ contains four scalar states $H$, $A$ and $H^\pm$. A discrete exact $\mathbb{Z}_2$ symmetry is introduced with the following transformation properties
\begin{equation}\label{eq:symm}
\phi_S\to \phi_S, \,\, \phi_D \to - \phi_D, \,\,
\text{SM} \to \text{SM}.
\end{equation}
The additional doublet does not acquire a vacuum expectation value (vev) and does not couple to fermions. Electroweak symmetry breaking  is as in the SM. The symmetry also insures that the lightest particle of  $\phi_D$ is stable, making this a good dark matter candidate.

The scalar potential of the model is given by
\begin{equation}\begin{array}{c}
V=-\frac{1}{2}\left[m_{11}^2(\phi_S^\dagger\phi_S)\!+\! m_{22}^2(\phi_D^\dagger\phi_D)\right]+
\frac{\lambda_1}{2}(\phi_S^\dagger\phi_S)^2\! 
+\!\frac{\lambda_2}{2}(\phi_D^\dagger\phi_D)^2\\[2mm]+\!\lambda_3(\phi_S^\dagger\phi_S)(\phi_D^\dagger\phi_D)\!
\!+\!\lambda_4(\phi_S^\dagger\phi_D)(\phi_D^\dagger\phi_S) +\frac{\lambda_5}{2}\left[(\phi_S^\dagger\phi_D)^2\!
+\!(\phi_D^\dagger\phi_S)^2\right].
\end{array}\label{pot}\end{equation}
The model features 7 free parameters, which we chose in the so-called physical basis {\cite{Ilnicka:2015jba}}
\begin{\eqn}\label{eq:physbas}
v,M_h,M_H, M_A, M_{H^{\pm}}, \lam_2, \lam_{345},
\end{\eqn}
with $\lam_{345}\,\equiv\,\lam_3+\lam_4+\lam_5$.
As two parameters (vev $v$ and $M_h\,\sim\,125\,\GeV$) are fixed by experimental measurements, we end up with a total number of 5 free parameters. Here, we consider the case where $H$ is the dark matter candidate, which implies $M_{A,\,H^\pm}\,\geq\,M_H$. \footnote{Note that the new scalars in the IDM do not have CP quantum numbers, as they do not couple to fermions. In the subsequent discussion, we can replace $H\,\longleftrightarrow\,A$ if we simultaneously use $\lam_5\,\longleftrightarrow\,-\lam_5$. All phenomenological considerations are identical for these cases.}

The model is subject to a large number of theoretical and experimental constraints, discussed at length e.g. in \cite{Ilnicka:2015jba,Ilnicka:2018def,Dercks:2018wch,Kalinowski:2018ylg,Kalinowski:2020rmb}. In the scan, we make use of the publicly available tools \texttt{2HDMC} \cite{Eriksson:2009ws},  \HBv5.10.1 \cite{Bechtle:2008jh, Bechtle:2011sb, Bechtle:2013wla,Bechtle:2015pma,Bechtle:2020pkv}, \HSv2.6.2 \cite{Bechtle:2013xfa,Bechtle:2020uwn}, as well as \texttt{micrOMEGAs$\_$5.2.4} \cite{Belanger:2020gnr}. Cross sections are calculated using  {\texttt{Madgraph5}} \cite{Alwall:2011uj} with a UFO input file from \cite{Goudelis:2013uca}\footnote{\label{foot:ufo} Note the official version available at \cite{ufo_idm} exhibits a wrong CKM structure, leading to false results for processes involving electroweak gauge bosons radiated off quark lines. In our implementation, we corrected for this. Our implementation corresponds to the expressions available from \cite{Zyla:2020zbs}.}. Experimental values are taken from GFitter \cite{gfitter,Haller:2018nnx}, and the Planck \cite{Aghanim:2018eyx} and XENON1T \cite{Aprile:2018dbl} experiments. Direct collider searches as well as agreement with the 125 \GeV coupling strength measurements are implemented via \HB~ and \HS, where we also compare to the total width upper limit  \cite{Sirunyan:2019twz} and invisible branching ratio \cite{ATLAS-CONF-2020-052} of $h$. Finally, recast results from a LEP-SUSY search \cite{Lundstrom:2008ai} were included.

\subsection{Current Status}
\begin{figure}[htb]
\begin{center}
\includegraphics[width=0.48\textwidth]{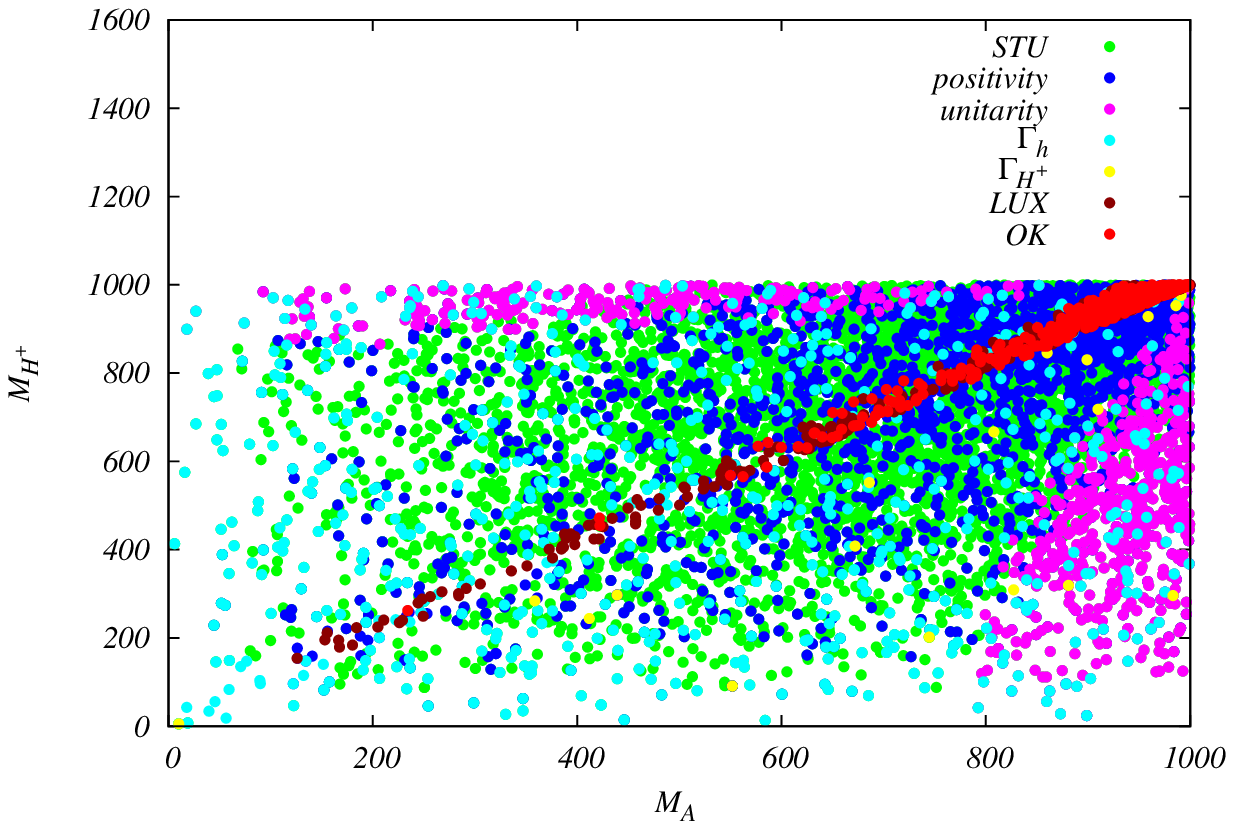}
\includegraphics[width=0.48\textwidth]{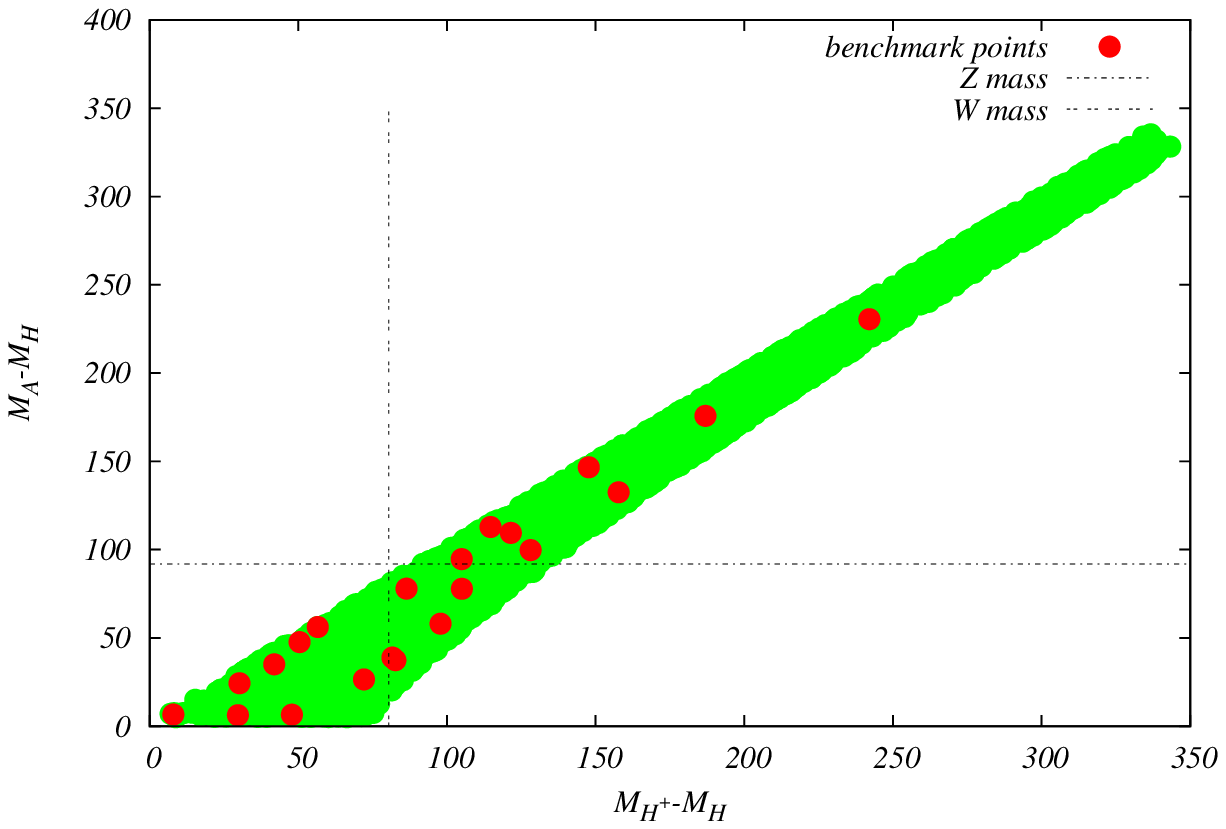}
\caption{Masses are requested to be quite degenerate after all constraints have been taken into account. {\sl Left:} In the $\lb M_A,\,M_{H^\pm} \rb$ plane (taken from \cite{Ilnicka:2015jba}). {\sl Right:} In the {$\lb M_{H^\pm}-M_H,\,M_A-M_H \rb$} plane (taken from \cite{Kalinowski:2018ylg}).}
\label{fig:massesidm}
\end{center}
\end{figure}

\begin{figure}[htb]
\centerline{%
\includegraphics[width=0.48\textwidth]{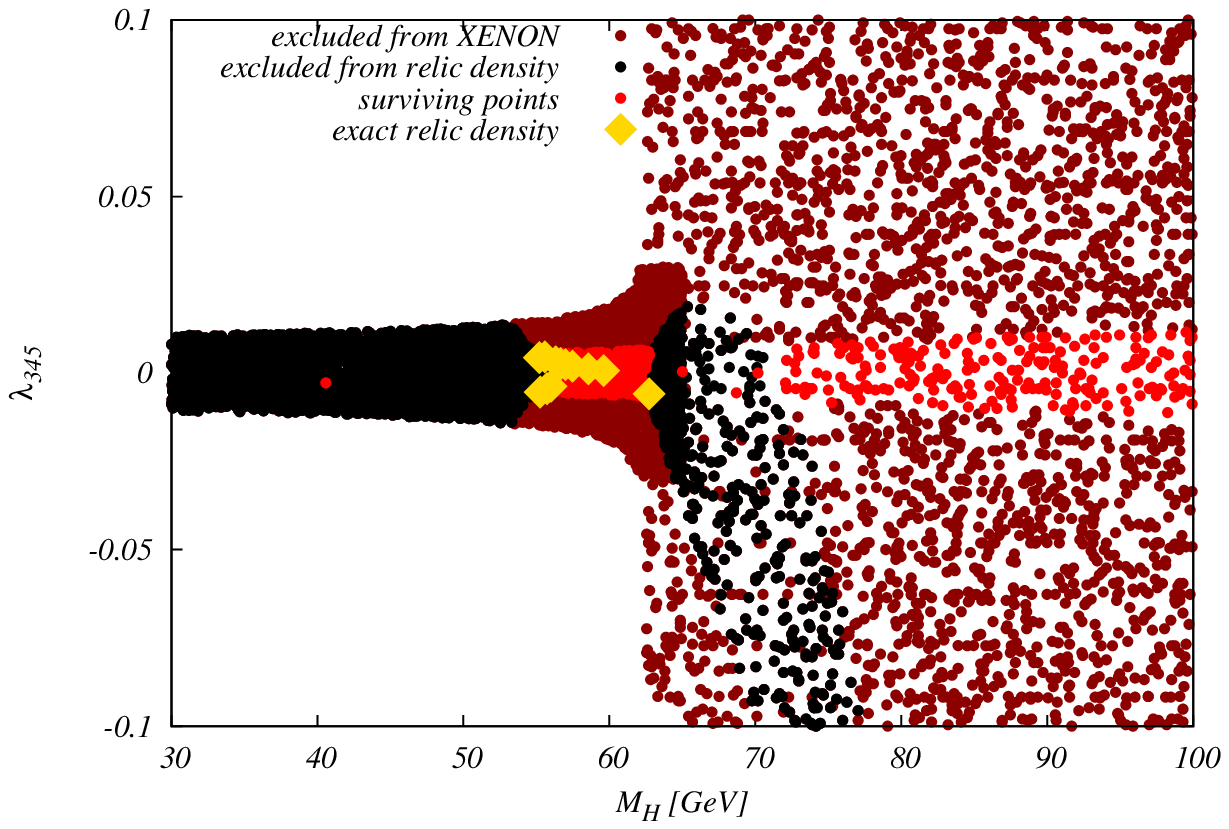}
\includegraphics[width=0.48\textwidth]{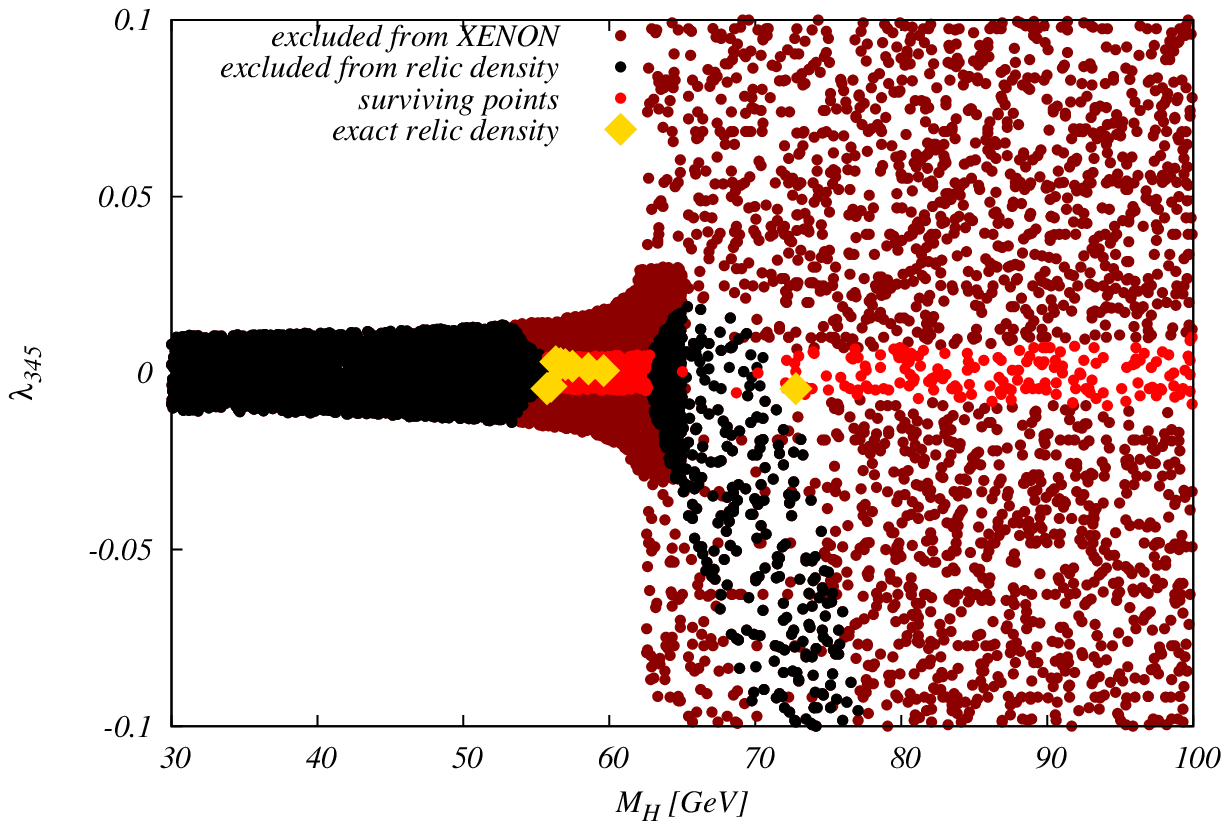}}
\caption{Interplay of signal strength and relic density constraints in the $\lb M_H,\,\lam_{345}\rb$ plane. {\sl Left:} Using LUX constraints \cite{Akerib:2013tjd}, bounds labelled "excluded from collider data" have been tested using \HB~and \HS~(taken from \cite{Ilnicka:2015jba}). {\sl Right:} Using XENON1T results, with golden points labelling those points that produce exact relic density (taken from \cite{Ilnicka:2018def}).}
\label{fig:lowmh}
\end{figure}

The experimental and theoretical constraints lead to a large reduction of the allowed parameter space. As an example, the masses are usually quite degenerate, as can be seen {from} figure \ref{fig:massesidm}. This is caused by an interplay of electroweak constraints and theoretical requirements on the potential. 
We also consider the case when $M_H\,\leq\,M_h/2$, where constraints from  $h\,\rightarrow\,\text{inv{isible}}$ start to play an important role and an interesting interplay arises, between bounds from signal strength measurements, that require $|\lam_{345}|$ to be rather small $\lesssim\,0.3$, and bounds from dark matter relic density, where too low values of that parameter lead to small annihilation cross sections and therefore too large relic density values. The resulting parameter space is shown in figure \ref{fig:lowmh}. In \cite{Ilnicka:2015jba}, it was found that this in general leads to a lower bound of $M_H\,\sim\,50\,\GeV$, with exceptions presented in \cite{Kalinowski:2020rmb}. 
\subsection{Discovery prospects at ILC and CLIC}
So far, no publicly available search exists that investigates the IDM parameter space with actual collider data. The discovery potential of ILC and CLIC was investigated in \cite{Kalinowski:2018kdn,deBlas:2018mhx,Zarnecki:2019poj,Zarnecki:2020swm,Sokolowska:2019xhe,Klamka:2022ukx} for several benchmark points proposed in \cite{Kalinowski:2018ylg}, for varying center-of-mass energies from $250\,\GeV$ up to $3\,\TeV$. We concentrated on $AH$ and $H^+ H^-$ production with $A\,\rightarrow\,Z\,H$ and $H^\pm\,\rightarrow\,W^\pm H$, where the electroweak gauge bosons subsequently decay leptonically. For event generation, we used \texttt{WHizard 2.2.8} \cite{Moretti:2001zz,Kilian:2007gr}, with an interface via \texttt{SARAH} \cite{Staub:2015kfa} and \texttt{SPheno 4.0.3} \cite{Porod:2003um,Porod:2011nf} for model implementation. For CLIC results  energy spectra \cite{Linssen:2012hp} were also taken into account.

\begin{figure}[htb]
\hspace{0.03\textwidth}
  \includegraphics[width=0.47\textwidth]{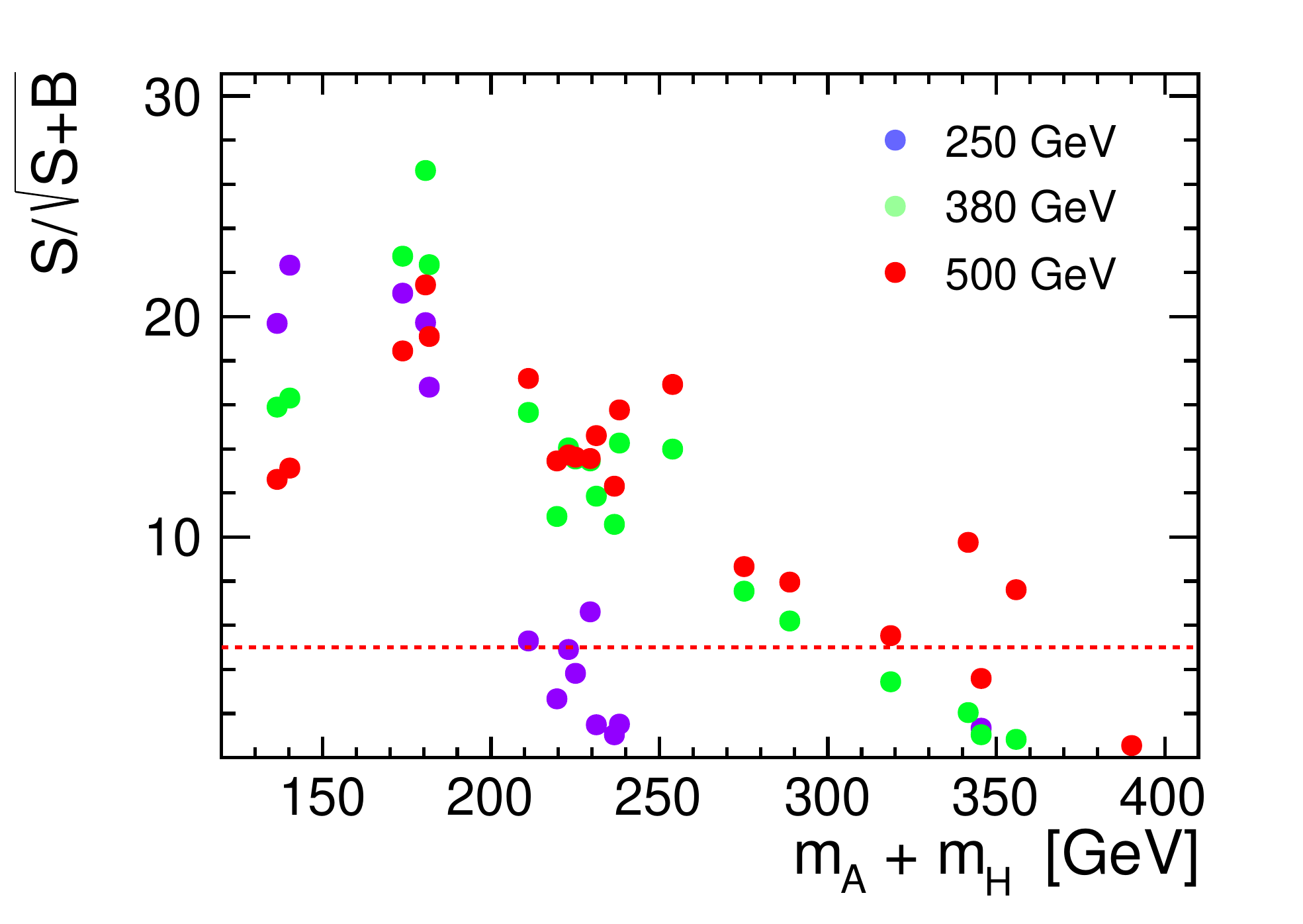}
  \includegraphics[width=0.47\textwidth]{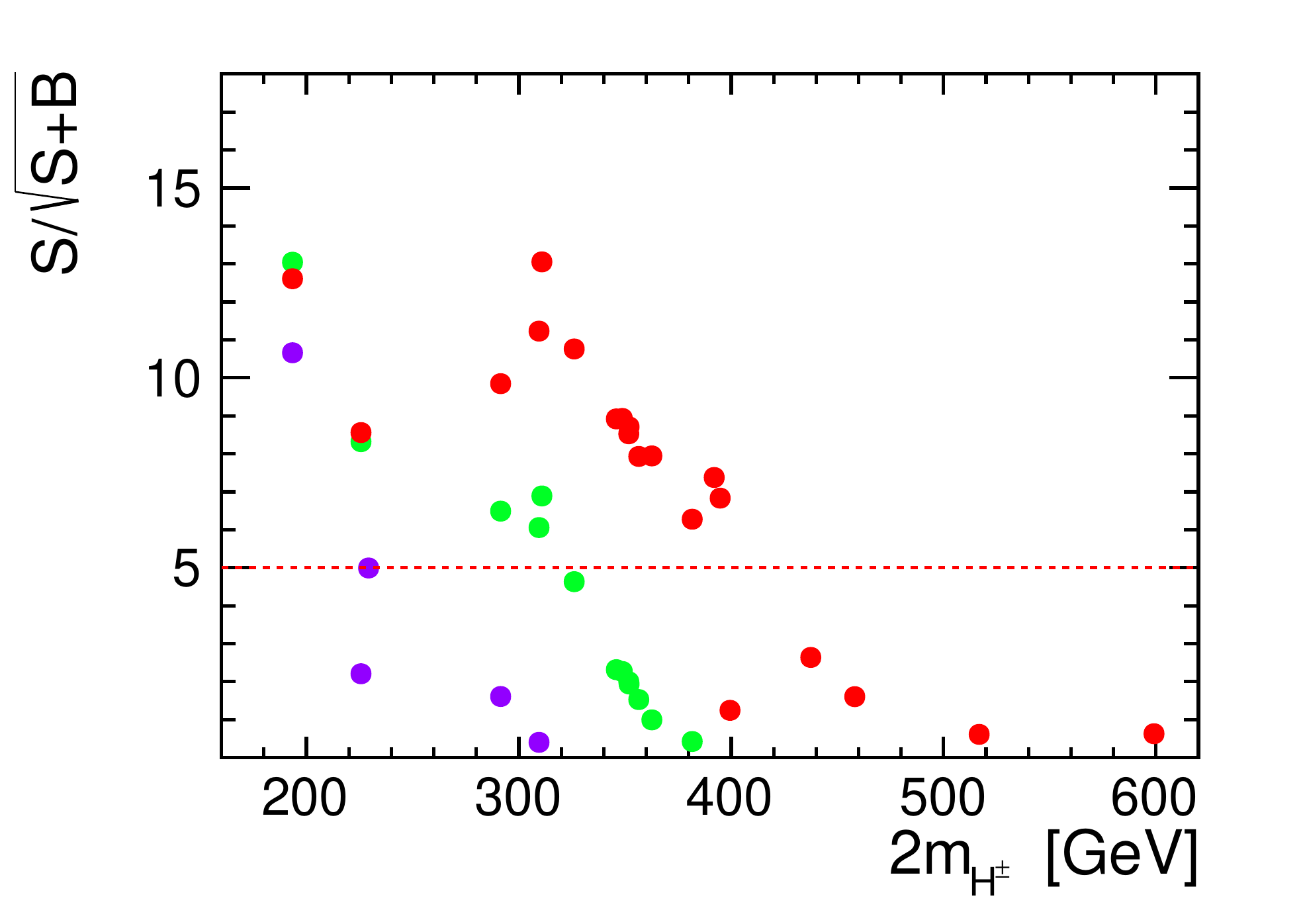}
\caption{Significance of the deviations from the Standard Model
  predictions expected for 1\,ab$^{-1}$ of data collected at
  centre-of-mass energy of 250\,GeV, 380\,GeV and   500\,GeV, for:
  (left) events with two muons in the final state ($\mu^+\mu^-$)
  as a function of the sum of neutral inert scalar masses and
  (right) events with an electron and a muon in the final state
  ($e^+\mu^-$ or $e^-\mu^+$) as a function of twice the charged scalar
  mass.  
}\label{fig:ilc}
\end{figure}

\begin{figure}[htb]
\begin{center}
\includegraphics[width=0.48\textwidth]{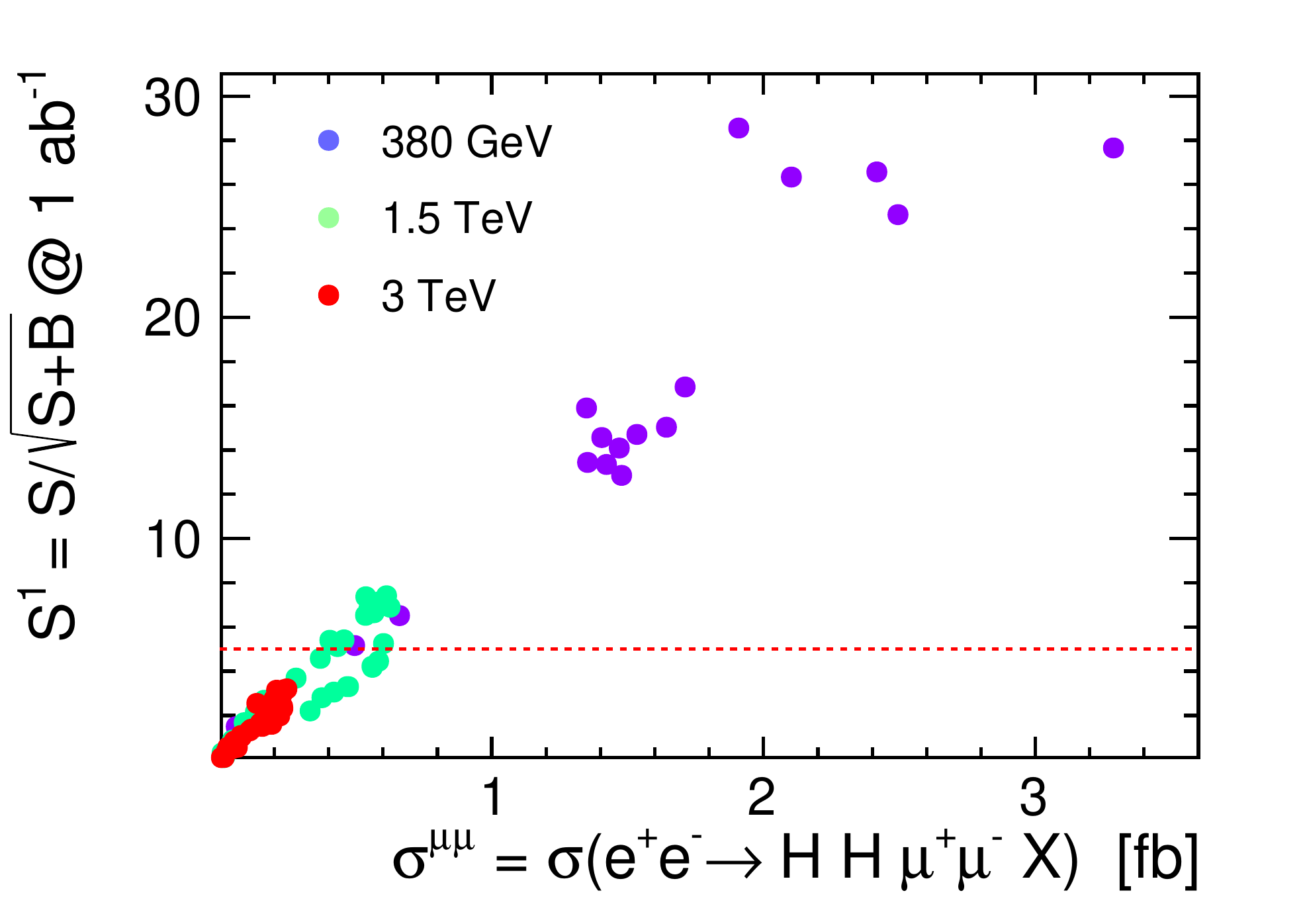}
\includegraphics[width=0.48\textwidth]{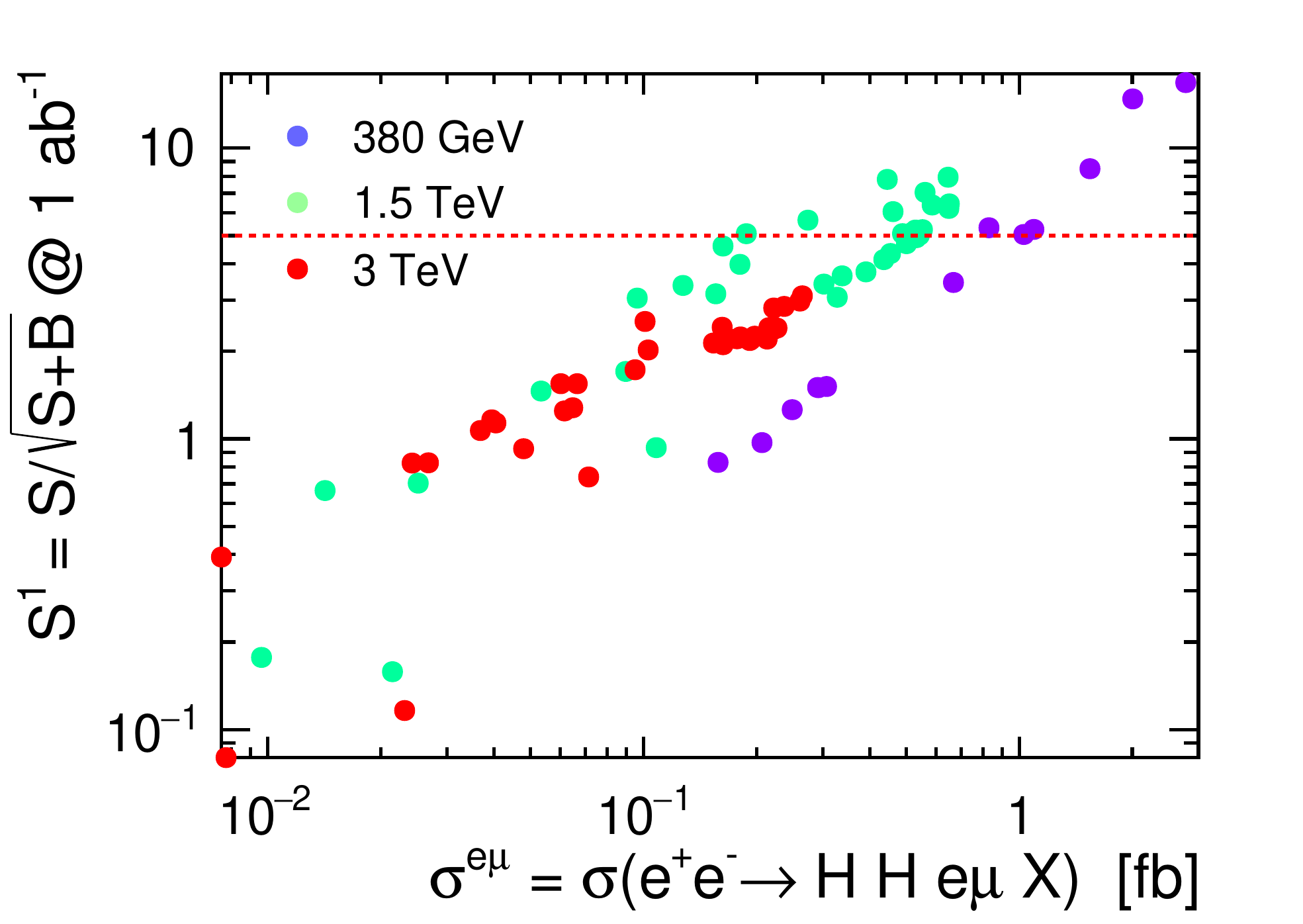}
\includegraphics[width=0.48\textwidth]{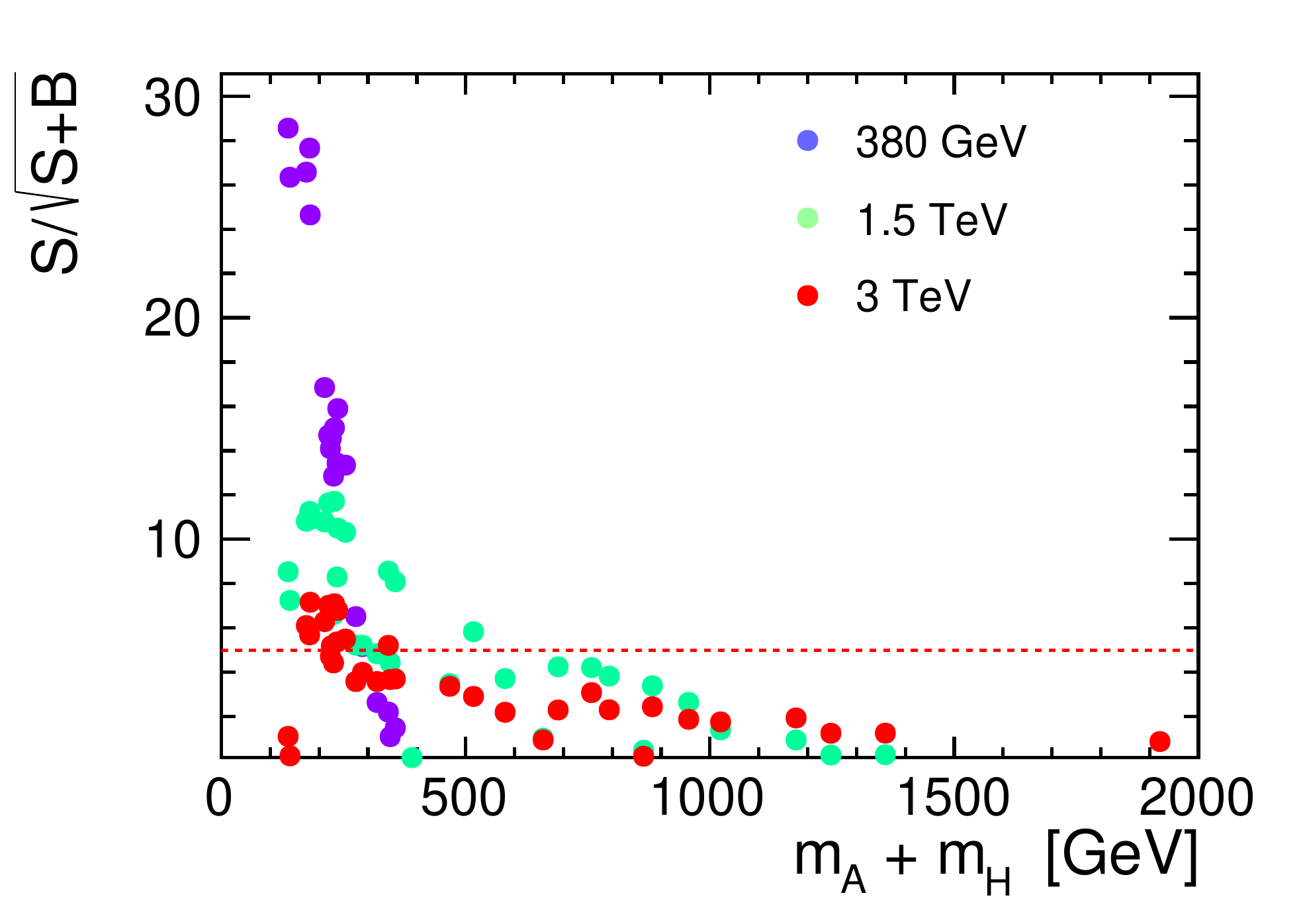}
\includegraphics[width=0.48\textwidth]{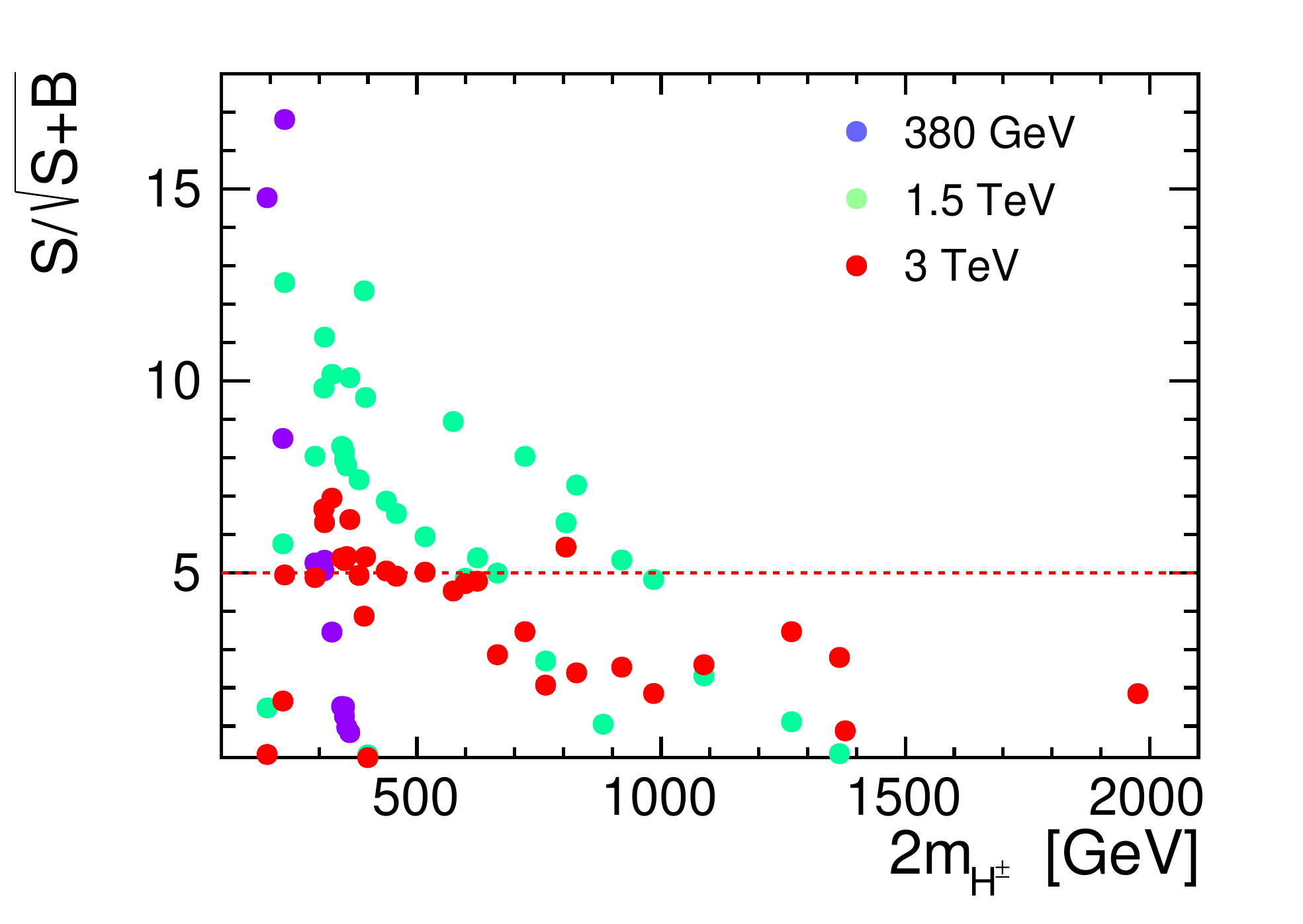}
\end{center}
\caption{Discovery prospects at CLIC for the IDM in $\mu^+\mu^-+\slashed{E}$ {\sl (left)} and $\mu^\pm\,e^\mp+\slashed{E}$ {\sl (right)} final states, as a function of the respective production cross-sections {\sl (top)} and mass sum of the produced particles {\sl (bottom)}. Taken from \cite{Kalinowski:2018kdn}.}
\label{fig:clic}
\end{figure}

For the production modes above, we considered leptonic decays of the electroweak gauge bosons. The investigated final states were
\begin{\eqn*}
e^+\,e^-\,\rightarrow\,\mu^+\mu^-+\slashed{E},\;\;\;
e^+\,e^-\,\rightarrow\,\mu^\pm\,e^\mp+\slashed{E}
\end{\eqn*}
for $HA$ and $H^+\,H^-$ production, respectively. For a more accurate description, we did not specify the intermediate states in the event generation, which in turn means all processes leading to the above signatures were taken into account, including interference between the contributing diagrams. This includes final states, where the missing energy stems from additional neutrinos, $e.g.$ from $\tau^\pm$ decays. 
Event selection was performed using a set of preselection cuts as well as boosted decision trees, as implemented in the TMVA toolkit \cite{Hocker:2007ht}. 
Results for the ILC running at 250\,GeV and 500\,GeV, and the first CLIC stage at 380\,GeV are shown in figure \ref{fig:ilc}. The expected discovery reach of 500\,GeV ILC extends up to neutral scalar mass sum of 330 GeV up to charged scalar masses of 200 GeV. 
Results for the discovery reach of CLIC, including center-of-mass energies of 1.5\,TeV and 3\,TeV, are shown in figure \ref{fig:clic}. In general, production cross sections $\gtrsim\,0.5\,\fb$ seem to be accessible, where best prospects for the considered benchmark points are given for 380 \GeV or 1.5 \TeV center-of-mass energies. Along similar lines, mass sums up to 1 \TeV seem accessible, where the $\mu^\pm\,e^\mp$ channel seems to provide a larger discovery range in general.

\begin{figure}[htb]
\begin{center}
\includegraphics[width=0.48\textwidth]{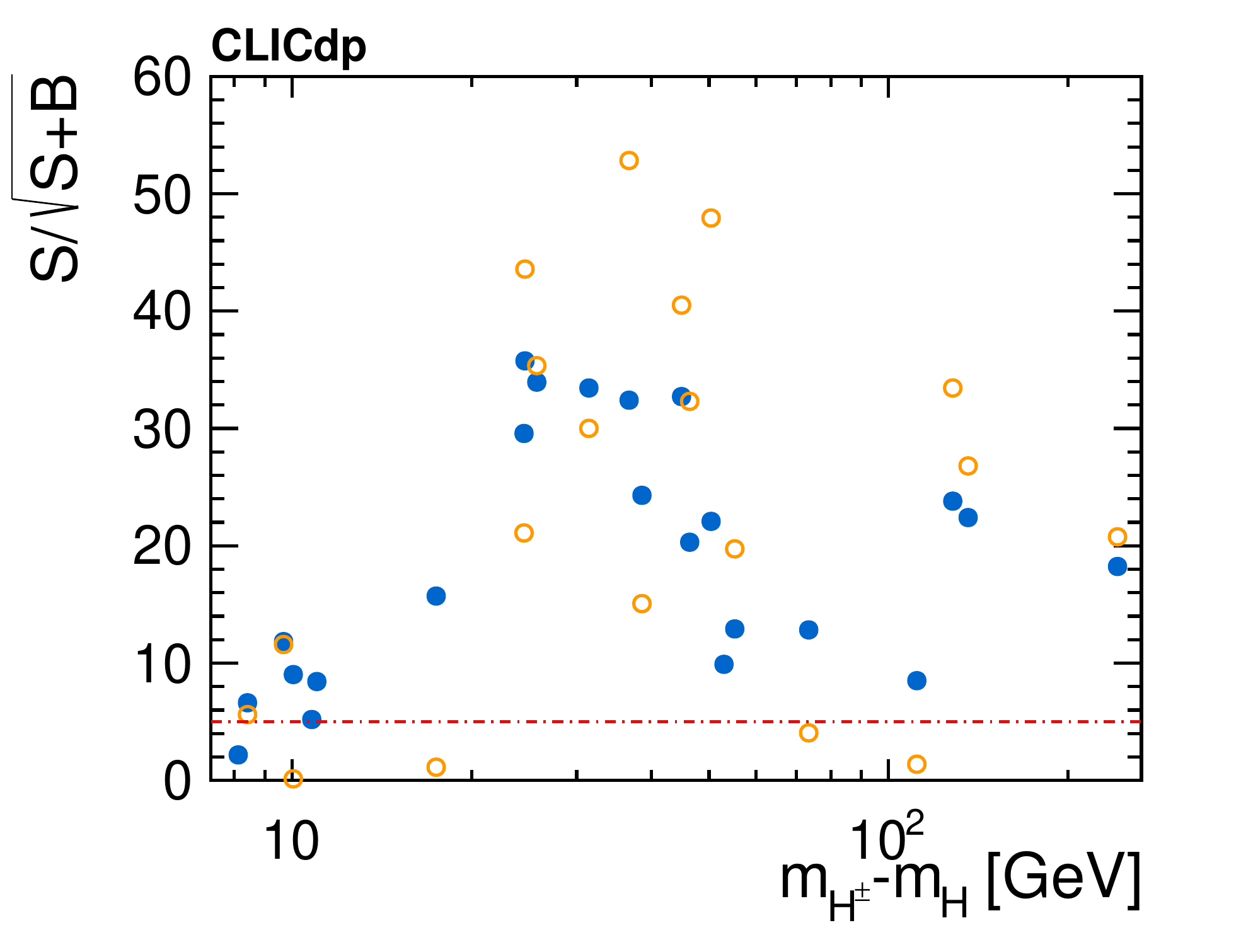}
\includegraphics[width=0.48\textwidth]{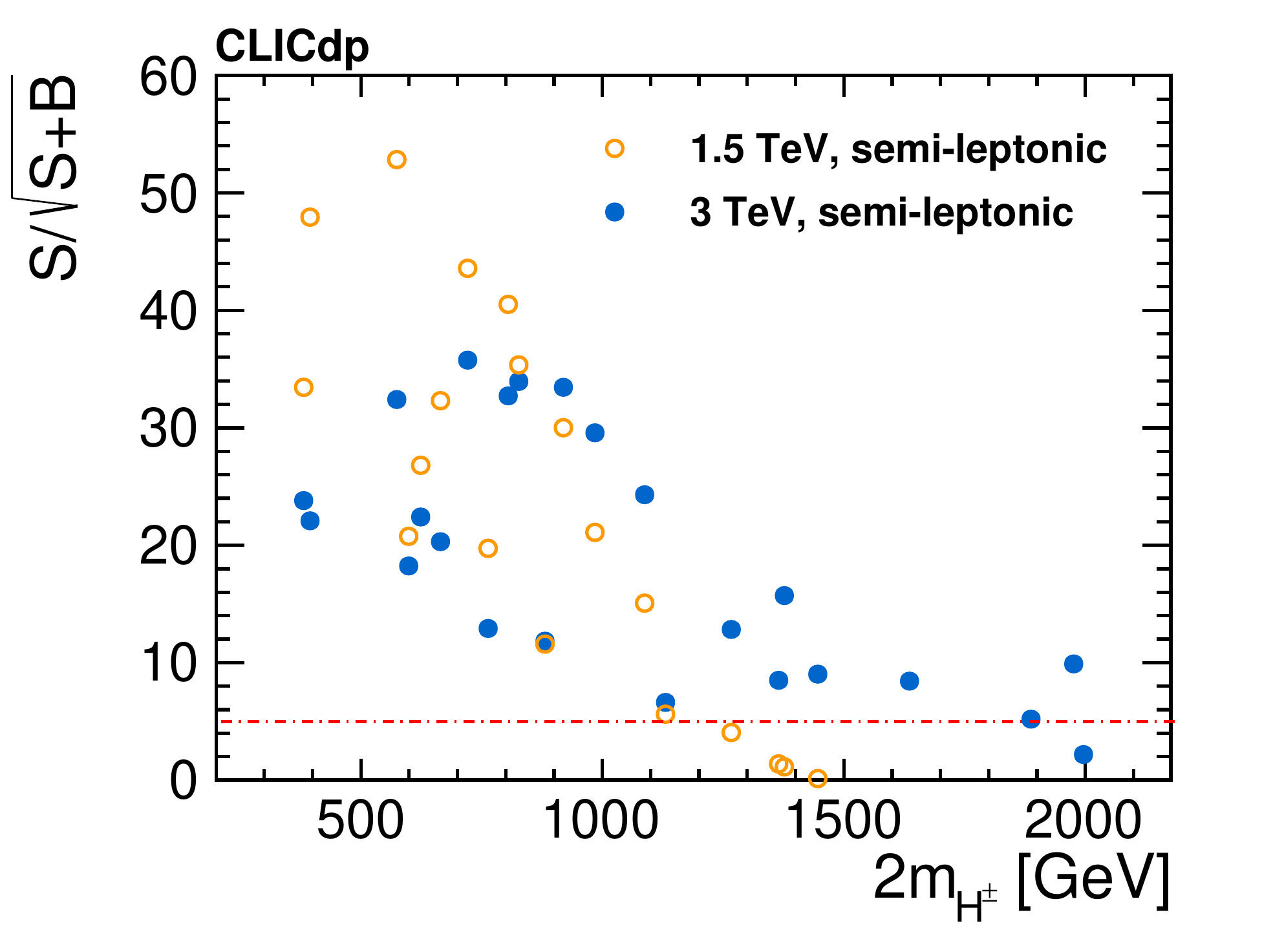}
\end{center}
\caption{Expected statistical significance of IDM charged scalar
  pair-production observation as a function of the IDM scalar
  mass difference (left) and of the
  total mass of the produced IDM scalars (right).
  Results are presented for CLIC running at 1.5\,TeV (orange circles) and 3\,TeV
  (blue points). The red horizontal lines indicate the 
  5$\sigma$ threshold. Figure taken from \cite{Klamka:2022ukx}.} 
	 	 \label{fig:clicsl}
	 \end{figure}

For leptonic signatures, the sensitivity of high energy $e^+e^-$ colliders to pair-production of IDM scalars is limited by the production cross section in the considered channel. 
The signal cross section for the charged scalar pair-production increases by about an order of magnitude when the semi-leptonic final state is considered,  i.e. when the  hadronic decay of one of the $W$ bosons is considered. %
The sensitivity of high energy CLIC to $H^+\,H^-$ production with the semi-leptonic final state is shown in Fig.~\ref{fig:clicsl}.
The accessible scalar mass range increases by about a factor of two, to about 2\,TeV \cite{Sokolowska:2019xhe,Zarnecki:2020swm,Klamka:2022ukx}.
It is also interesting to note that highest signal observation significance is obtained for scenarios with scalar mass difference of 20 to 60 GeV. 

\subsection{Sensitivity comparison at future colliders}
After a dedicated analysis of the IDM benchmarks in the high energy CLIC environment, an important question is whether other current or future collider options provide similar or better discovery prospects. For the same set of benchmarks\cite{Kalinowski:2018ylg,Kalinowski:2018kdn}, production cross sections for a variety of processes have been presented in \cite{Kalinowski:2020rmb}, including VBF-type topologies. Cross sections were calculated using {\texttt{Madgraph5}}. 
We concentrate on production modes at muon colliders and refer to \cite{Kalinowski:2020rmb} regarding processes at proton-proton machines. We list the considered collider types and nominal center-of-mass energies as well as integrated luminosities in table \ref{tab:colls}.
\begin{center}
\begin{table}
\begin{center}
\begin{tabular}{c|c|c|c}
collider&cm energy [\TeV]&$\int\mathcal{L}$&{$\sigma_{_{1000}}$} [\fb]\\ \hline
ee&3&$5\,\ab^{-1}$&0.2\\
$\mu\mu$&10&$10\,\ab^{-1}$&0.1\\
$\mu\mu$&30&$90\,\ab^{-1}$&0.01
\end{tabular}
\end{center}
\caption{Collider parameters used in the discovery reach {study} performed in \cite{Kalinowski:2020rmb}. Collider specifications have been taken from \cite{Delahaye:2019omf} for the muon collider. The last column denotes the minimal cross section {required to produce} 1000 events using full target luminosity.  } 
\label{tab:colls}
\end{table}
\end{center}

A scenario is called "realistic" when  1000 events can be produced using target luminosity and center-of-mass energies as specified above. Obviously, more detailed studies, including {both} background {contribution and detector response} simulation, are necessary to assess the actual collider reach.

We here concentrate on production at future muon colliders:
\begin{\eqn*}
  \mu^+\,\mu^-\,\rightarrow\,\nu_\mu\,\bar{\nu}_\mu A A,\;\;\;
  \mu^+\,\mu^-\,\rightarrow\,\nu_\mu\,\bar{\nu}_\mu H^+ H^-.
\end{\eqn*}
which corresponds to VBF-like production modes. Here again intermediate states are not specified, so in fact {several} diagrams contribute which not all have a typical VBF topology. See appendix B and C of \cite{Kalinowski:2020rmb} for details.

\begin{figure}[htb]
\begin{center}
\includegraphics[width=0.48\textwidth]{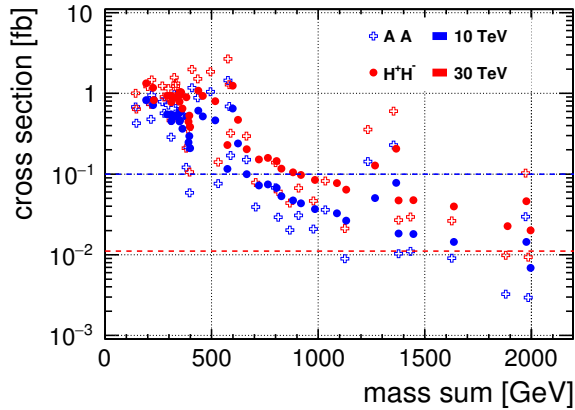}
\end{center}
\caption{Cross sections as a function of the mass sum at muon colliders in the VBF-type production mode.  Taken from \cite{Kalinowski:2020rmb}.}
\label{fig:vbf}
\end{figure}

Figure \ref{fig:vbf} shows the production cross sections as a function of the mass sum of produced particles for various center-of-mass energies. Understanding the behaviour of the VBF-induced channels {is} non-trivial; this can be attributed to the fact that more diagrams, apart from the naive pair-production process, contribute. Large jumps between cross-section predictions for scenarios with similar mass scales can be traced back mainly to a fine-tuned cancellation of various contributing diagrams, as discussed in greater  detail in \cite{Kalinowski:2020rmb}.

\begin{center}
\begin{table}
\begin{center}
\begin{tabular}{||c||c||c||c||} \hline \hline
{collider}&{all others}& { $AA$} & {$AA$ +VBF}\\ \hline \hline
HL-LHC&1 \TeV&200-600 \GeV& 500-600 \GeV\\
HE-LHC&2 \TeV&400-1400 \GeV&800-1400 \GeV\\
FCC-hh&2 \TeV&600-2000 \GeV&1600-2000 \GeV\\ \hline \hline
CLIC, 3 \TeV&2 \TeV &- &300-600 \GeV\\
$\mu\mu$, 10 \TeV&2 \TeV &-&400-1400 \GeV\\
$\mu\mu$, 30 \TeV&2 \TeV  &-&1800-2000 \GeV \\ \hline \hline
\end{tabular}
\end{center}
\caption{Sensitivity of different collider options specified in table \ref{tab:colls}, using the "realistic" criterium of 1000 generated events in the specific channel. Shown are  minimal and  maximal mass scales that are reachable. Numbers for CLIC correspond to results from detailed investigations \cite{Kalinowski:2018kdn,deBlas:2018mhx}. Table taken from \cite{Robens:2021zvr}. }
\label{tab:sens}
\end{table}
\end{center}
The summary of sensitivities in terms of mass scales is given in table \ref{tab:sens}, where for completeness the reach  for proton colliders as  estimated in the above work has been added.  It is  seen that especially for $AA$ production the VBF mode at both proton and muon  colliders serves to significantly increase the discovery reach of the respective machine. Using the simple counting criterium above, we can furthermore state that a 27 \TeV\ proton-proton machine has a similar reach as a 10 \TeV\ muon collider, while 100 \TeV\ FCC-hh would correspond to a 30 \TeV\ muon-muon machine. Obviously, detailed investigations including SM background are needed to give a more realistic estimate of the respective collider reach.
\\

\section{THDMa}
The THDMa is a type II two-Higgs-doublet model that is extended by an additional pseudoscalar $a$ mixing with the "standard" pseudoscalar $A$ of the THDM. In the gauge-eigenbasis, the additional scalar serves as a portal to the dark sector, with a fermionic dark matter candidate, denoted by $\chi$. More details can e.g. be found in \cite{Ipek:2014gua,No:2015xqa,Goncalves:2016iyg,Bauer:2017ota,Tunney:2017yfp,LHCDarkMatterWorkingGroup:2018ufk,Robens:2021lov}. 

The model contains the following particles in the scalar and dark matter sector: ${h,\,H,\,H^\pm}$, ${a,}\,{A,}\,{{\chi}}$. It depends on 12 additional new physics parameters
\begin{eqnarray*}
{v,\,m_h,\,m_H,}\,{ m_a,}\,{m_A,\,m_{H^\pm},}\,{m_\chi};\;{\cos\lb \be-\al\rb,\,\tan\be,}\,{\sin\theta;\;y_\chi,}\,{\lam_3,}\,{\lam_{P_1},\,\lam_{P_2}},
\end{eqnarray*}
where $v$ and either $m_h$ or $m_H$ are fixed by current measurements in the electroweak sector. 

\begin{center}
\begin{figure}
\begin{center}
\begin{minipage}{0.45\textwidth}
\begin{center}
\includegraphics[width=\textwidth]{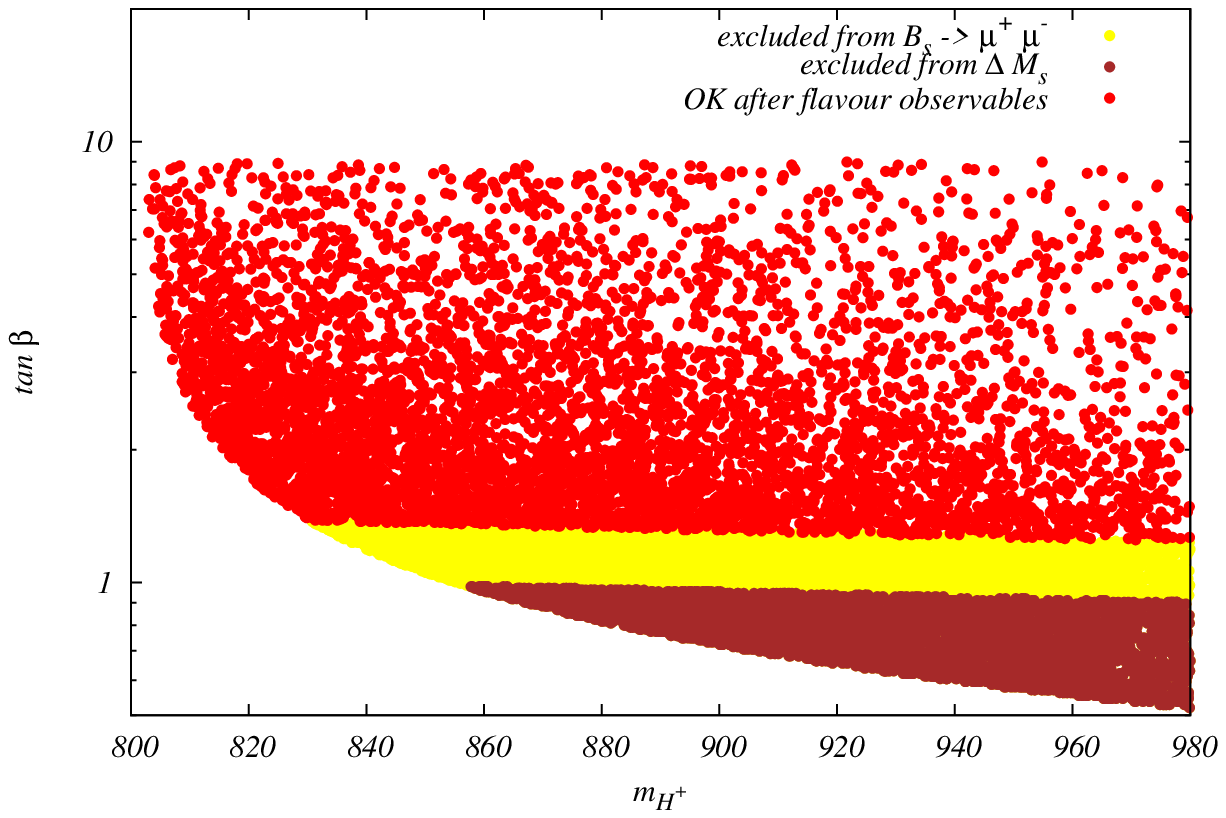}
\end{center}
\end{minipage}
\begin{minipage}{0.45\textwidth}
\begin{center}
\includegraphics[width=\textwidth]{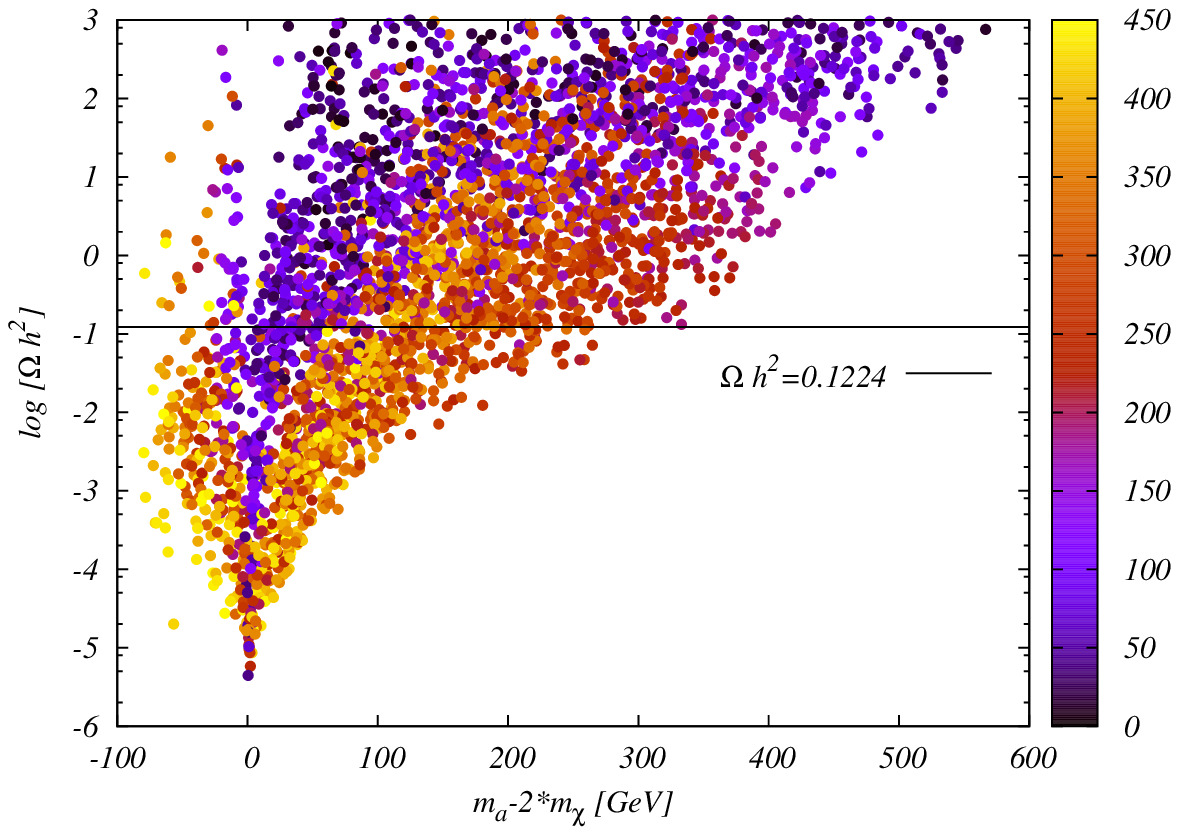}
\end{center}
\end{minipage}
\end{center}
\caption{\label{fig:thdmab} {\sl Left:} Bounds on the $\lb m_{H^\pm},\,\tan\be\rb$ plane from B-physics observables, implemented via the SPheno \cite{Porod:2011nf}/ Sarah \cite{Staub:2013tta} interface,  and compared to experimental bounds \cite{combi,Amhis:2019ckw}. The contour for low $\lb m_{H^\pm,\,\tan\be}\rb$ values stems from \cite{Misiak:2020vlo,mm}. {\sl Right:} Dark matter constraints in the THDMa model. {\sl Right:} Dark matter relic density as a function of $m_a-2\,m_\chi$, with $m_\chi$ defining the color coding. The typical resonance-enhanced relic density annihilation is clearly visible. Figures taken from \cite{Robens:2021lov}.}
\end{figure}
\end{center}

We here report on results of a scan that allows all of the above novel parameters float in specific predefined ranges \cite{Robens:2021lov}. In such a scenario, it is not always straightforward to display bounds from specific constraints in 2-dimensional planes. Two examples for scenarios where this is possible are shown in figure \ref{fig:thdmab}. The first plot shows bounds in the $\lb m_{H^\pm},\,\tan\be \rb$ plane from B-physics observables. The result is similar to a simple THDM, and shows that in general low masses $m_{H^\pm}\lesssim\,800\,\GeV$  as well as values $\tan\be\lesssim\,1$ are excluded. The second plot displays the relic density as a function of the mass difference $m_a-2\,m_\chi$. Here, a behaviour can be observed that is typical in many models with dark matter candidates: in the region where this mass difference remains small, relic density annihilates sufficiently to stay below the observed relic density bound. On the other hand, too large differences lead to values $\Omega\,h_c\,\gtrsim\,0.12$ and therefore are forbidden from dark matter considerations.

\begin{center}
\begin{figure}
\includegraphics[width=0.67\textwidth]{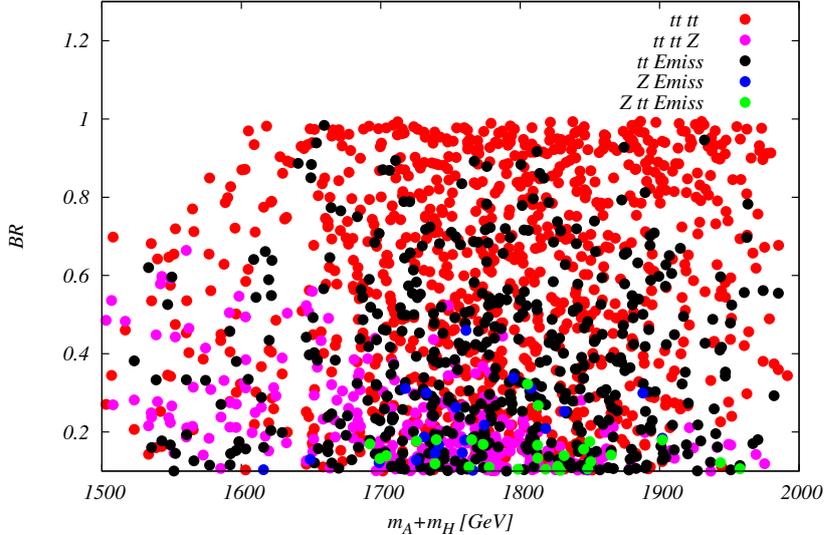}
\caption{\label{fig:brsAH} Branching ratios into various final states for $A\,H$ production, as a function of the mass sum.}
\end{figure}
\end{center}

\begin{center}
\begin{figure}
\begin{center}
\begin{minipage}{0.45\textwidth}
\begin{center}
\includegraphics[width=\textwidth]{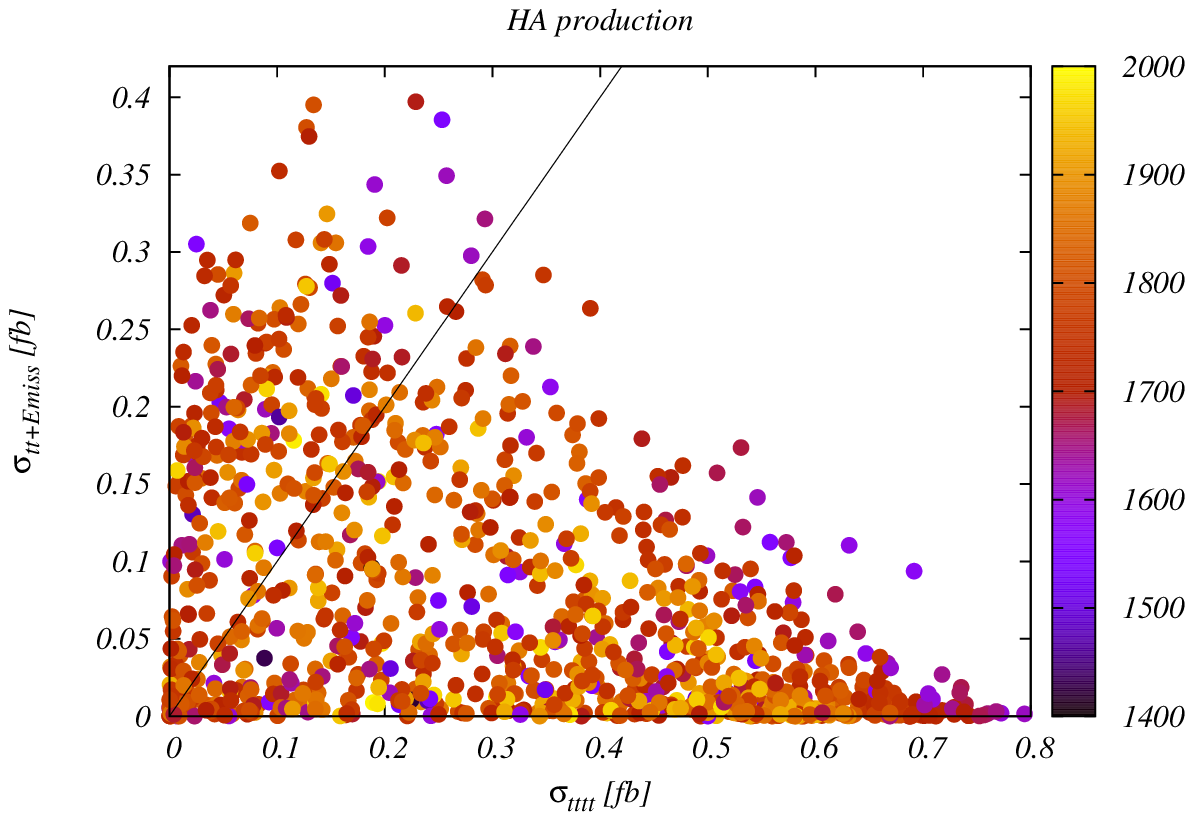}
\end{center}
\end{minipage}
\begin{minipage}{0.45\textwidth}
\begin{center}
\includegraphics[width=\textwidth]{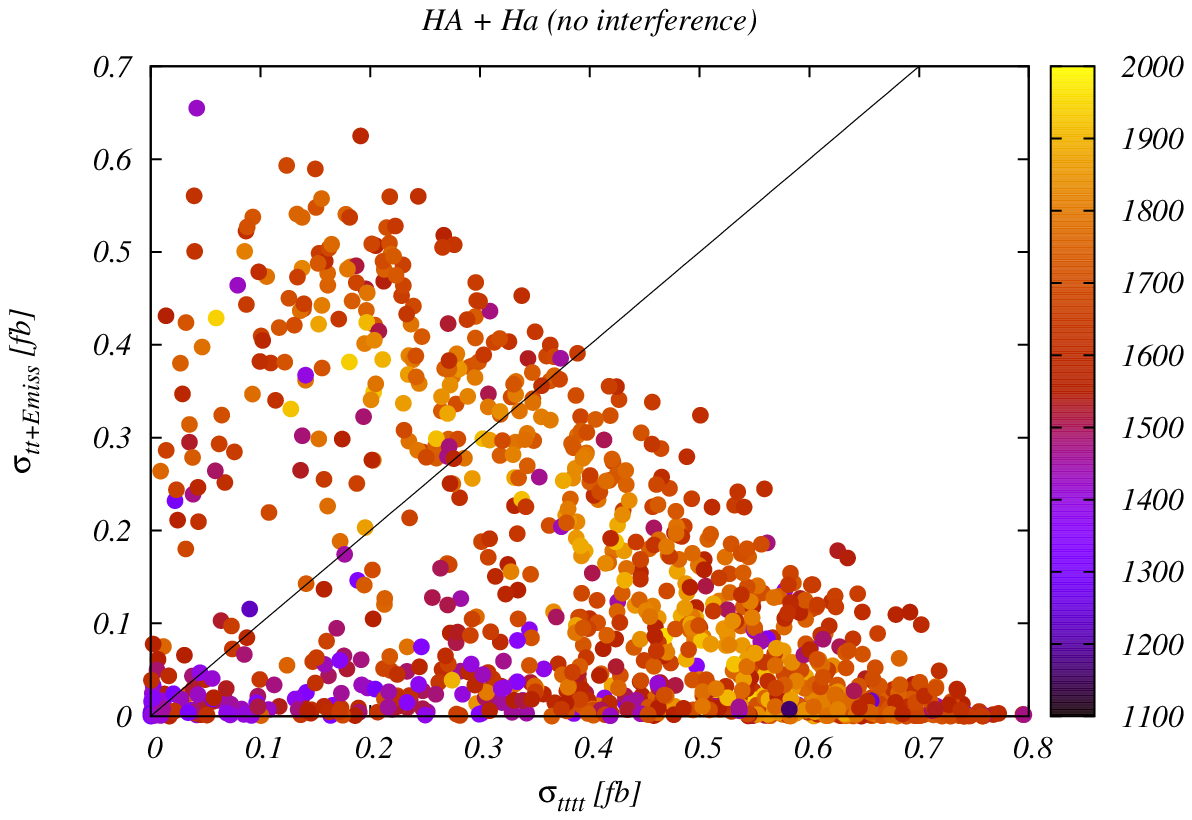}
\end{center}
\end{minipage}
\end{center}
\caption{\label{fig:thdmaatee} Production cross sections for $t\bar{t}t\bar{t}$ (x-axis) and $t\bar{t}+\slashed{E}$ (y-axis) final state in a factorized approach, for an $e^+e^-$ collider with a 3 \TeV center-of-mass energy. {\sl Left:} mediated via $HA$, {\sl right:} mediated via $HA$ and $Ha$ intermediate states. Color coding refers to $m_H+m_A$ {\sl (left)} and $M_H+0.5\times\,\lb m_A+m_a\rb$ {\sl(right)}. Figures taken from \cite{Robens:2021lov}.}
\end{figure}
\end{center}

Finally, it was investigated which cross-section values would still be feasible for points that fulfill all constraints \cite{Robens:2021lov} at $e^+e^-$ colliders. We  here concentrate on signatures that include missing energy and therefore do not exist in a THDM without a portal to the dark sector. Processes like $e^+e^-\,\rightarrow\,hA, ha$ are suppressed due to alignment, which makes $e^+e^-\,\rightarrow\,HA, Ha$ the most interesting channel that contains novel signatures. Due to the interplay of B-physics and electroweak constraints, such points typically have mass scales $\gtrsim\,1\,\TeV$. Therefore production cross sections for an $e^+e^-$ collider with a center-of-mass energy of 3 \TeV\ are of interest. The corresponding production cross sections are shown in figure \ref{fig:thdmaatee}, which displays  predictions for $t\,\bar{t}\,t\,\bar{t}$ and $t\,\bar{t}+\slashed{E}$ final states using a factorized approach. There is a non-negligible number of points where the second channel is dominant. A "best" point with a large rate for $t\,\bar{t}+\slashed{E}_\perp$ has been presented in \cite{Robens:2021lov} and is repeated here for completeness

\begin{eqnarray}
&&\sin\theta\,=\,-0.626,\;\cos\lb \be-\al\rb\,=\,0.0027,\,\tan\be\,=\,3.55 \nonumber\\
&&m_H\,=\,643\,\GeV,\,m_A\,=\,907\,\GeV,\,m_{H^\pm}\,=\,814\,\GeV, \nonumber\\
&&m_a\,\,=\,653\,\GeV,\,m_\chi\,=\,277\,\GeV, \nonumber\\
&&y_\chi\,\,=\,-1.73,\,\lambda_{P_1}\,=\,0.18,\,\lambda_{P_2}\,=\,2.98,\,\lambda_3\,=\,8.63.
\end{eqnarray}
For this point, all width/ mass ratios are $\lesssim\,6\,\%$. In addition, branching ratios for various final states as a function of the mass sum for the $HA$ channel are given in figure \ref{fig:brsAH}.

\section{Conclusion and Outlook}
In this work, we presented two models that extend the particle content of the SM and also provide at least one dark matter candidate. We have presented production cross sections for various standard pair-production modes within these models; for the IDM, we have given an estimate of mass range that can be reached  based on a simple counting criterium. A more dedicated investigation of the corresponding signatures, including background simulation and cut optimization, is in the line of future work.
\section*{Acknowledgements}
This research was supported in parts by the National Science Centre, Poland, the HARMONIA
project under contract UMO-2015/18/M/ST2/00518 (2016-2021), OPUS project under contract
UMO-2017/25/B/ST2/00496 (2018-2021), as well as the COST action CA16201 - Particleface. 



\begin{thebibliography}{10}

\bibitem{Bechtle:2020pkv}
Philip Bechtle, Daniel Dercks, Sven Heinemeyer, Tobias Klingl, Tim Stefaniak,
  Georg Weiglein, and Jonas Wittbrodt.
\newblock {HiggsBounds-5: Testing Higgs Sectors in the LHC 13 TeV Era}.
\newblock {\em Eur. Phys. J.}, C80(12):1211, 2020, 2006.06007.

\bibitem{Bechtle:2020uwn}
Philip Bechtle, Sven Heinemeyer, Tobias Klingl, Tim Stefaniak, Georg Weiglein,
  and Jonas Wittbrodt.
\newblock {HiggsSignals-2: Probing new physics with precision Higgs
  measurements in the LHC 13 TeV era}.
\newblock {\em Eur. Phys. J. C}, 81(2):145, 2021, 2012.09197.

\bibitem{Eriksson:2009ws}
David Eriksson, Johan Rathsman, and Oscar St{\aa}l.
\newblock {2HDMC: Two-Higgs-Doublet Model Calculator Physics and Manual}.
\newblock {\em Comput. Phys. Commun.}, 181:189--205, 2010, 0902.0851.

\bibitem{Porod:2011nf}
W.~Porod and F.~Staub.
\newblock {SPheno 3.1: Extensions including flavour, CP-phases and models
  beyond the MSSM}.
\newblock {\em Comput. Phys. Commun.}, 183:2458--2469, 2012, 1104.1573.

\bibitem{Staub:2013tta}
Florian Staub.
\newblock {SARAH 4 : A tool for (not only SUSY) model builders}.
\newblock {\em Comput. Phys. Commun.}, 185:1773--1790, 2014, 1309.7223.

\bibitem{Belanger:2018ccd}
Genevi\`eve B\'elanger, Fawzi Boudjema, Andreas Goudelis, Alexander Pukhov, and
  Bryan Zaldivar.
\newblock {micrOMEGAs5.0 : Freeze-in}.
\newblock {\em Comput. Phys. Commun.}, 231:173--186, 2018, 1801.03509.

\bibitem{Belanger:2020gnr}
Genevieve Belanger, Ali Mjallal, and Alexander Pukhov.
\newblock {Recasting direct detection limits within micrOMEGAs and implication
  for non-standard Dark Matter scenarios}.
\newblock {\em Eur. Phys. J. C}, 81(3):239, 2021, 2003.08621.

\bibitem{Ambrogi:2018jqj}
Federico Ambrogi, Chiara Arina, Mihailo Backovic, Jan Heisig, Fabio Maltoni,
  Luca Mantani, Olivier Mattelaer, and Gopolang Mohlabeng.
\newblock {MadDM v.3.0: a Comprehensive Tool for Dark Matter Studies}.
\newblock {\em Phys. Dark Univ.}, 24:100249, 2019, 1804.00044.

\bibitem{Baak:2014ora}
M.~Baak, J.~Cúth, J.~Haller, A.~Hoecker, R.~Kogler, K.~Mönig, M.~Schott, and
  J.~Stelzer.
\newblock {The global electroweak fit at NNLO and prospects for the LHC and
  ILC}.
\newblock {\em Eur. Phys. J.}, C74:3046, 2014, 1407.3792.

\bibitem{Haller:2018nnx}
Johannes Haller, Andreas Hoecker, Roman Kogler, Klaus Moenig, Thomas Peiffer,
  and Joerg Stelzer.
\newblock {Update of the global electroweak fit and constraints on
  two-Higgs-doublet models}.
\newblock {\em Eur. Phys. J.}, C78(8):675, 2018, 1803.01853.

\bibitem{combi}
CMS-PAS-BPH-20-003, LHCb-CONF-2020-002, ATLAS-CONF-2020-049.

\bibitem{Amhis:2019ckw}
Yasmine~Sara Amhis et~al.
\newblock {Averages of b-hadron, c-hadron, and $\tau $-lepton properties as of
  2018}.
\newblock {\em Eur. Phys. J.}, C81(3):226, 2021, 1909.12524.

\bibitem{Planck:2018vyg}
N.~Aghanim et~al.
\newblock {Planck 2018 results. VI. Cosmological parameters}.
\newblock {\em Astron. Astrophys.}, 641:A6, 2020, 1807.06209.
\newblock [Erratum: Astron.Astrophys. 652, C4 (2021)].

\bibitem{Aprile:2018dbl}
E.~Aprile et~al.
\newblock {Dark Matter Search Results from a One Ton-Year Exposure of XENON1T}.
\newblock {\em Phys. Rev. Lett.}, 121(11):111302, 2018, 1805.12562.

\bibitem{Misiak:2020vlo}
M.~Misiak, Abdur Rehman, and Matthias Steinhauser.
\newblock {Towards $ \overline{B}\to {X}_s\gamma $ at the NNLO in QCD without
  interpolation in m$_{c}$}.
\newblock {\em JHEP}, 06:175, 2020, 2002.01548.

\bibitem{mm}
M.~Misiak.
\newblock Private communication.

\bibitem{Alwall:2011uj}
Johan Alwall, Michel Herquet, Fabio Maltoni, Olivier Mattelaer, and Tim
  Stelzer.
\newblock {MadGraph 5 : Going Beyond}.
\newblock {\em JHEP}, 06:128, 2011, 1106.0522.

\bibitem{Deshpande:1977rw}
Nilendra~G. Deshpande and Ernest Ma.
\newblock {Pattern of Symmetry Breaking with Two Higgs Doublets}.
\newblock {\em Phys. Rev.}, D18:2574, 1978.

\bibitem{Cao:2007rm}
Qing-Hong Cao, Ernest Ma, and G.~Rajasekaran.
\newblock {Observing the Dark Scalar Doublet and its Impact on the
  Standard-Model Higgs Boson at Colliders}.
\newblock {\em Phys. Rev.}, D76:095011, 2007, 0708.2939.

\bibitem{Barbieri:2006dq}
Riccardo Barbieri, Lawrence~J. Hall, and Vyacheslav~S. Rychkov.
\newblock {Improved naturalness with a heavy Higgs: An Alternative road to LHC
  physics}.
\newblock {\em Phys. Rev.}, D74:015007, 2006, hep-ph/0603188.

\bibitem{Ilnicka:2015jba}
Agnieszka Ilnicka, Maria Krawczyk, and Tania Robens.
\newblock {Inert Doublet Model in light of LHC Run I and astrophysical data}.
\newblock {\em Phys. Rev. D}, 93(5):055026, 2016, 1508.01671.

\bibitem{Ilnicka:2018def}
Agnieszka Ilnicka, Tania Robens, and Tim Stefaniak.
\newblock {Constraining Extended Scalar Sectors at the LHC and beyond}.
\newblock {\em Mod. Phys. Lett. A}, 33(10n11):1830007, 2018, 1803.03594.

\bibitem{Dercks:2018wch}
Daniel Dercks and Tania Robens.
\newblock {Constraining the Inert Doublet Model using Vector Boson Fusion}.
\newblock {\em Eur. Phys. J. C}, 79(11):924, 2019, 1812.07913.

\bibitem{Kalinowski:2018ylg}
Jan Kalinowski, Wojciech Kotlarski, Tania Robens, Dorota Sokolowska, and
  Aleksander~Filip Zarnecki.
\newblock {Benchmarking the Inert Doublet Model for $e^+ e^-$ colliders}.
\newblock {\em JHEP}, 12:081, 2018, 1809.07712.

\bibitem{Kalinowski:2020rmb}
Jan Kalinowski, Tania Robens, Dorota Sokolowska, and Aleksander~Filip Zarnecki.
\newblock {IDM Benchmarks for the LHC and Future Colliders}.
\newblock {\em Symmetry}, 13(6):991, 2021, 2012.14818.

\bibitem{Bechtle:2008jh}
Philip Bechtle, Oliver Brein, Sven Heinemeyer, Georg Weiglein, and Karina~E.
  Williams.
\newblock {HiggsBounds: Confronting Arbitrary Higgs Sectors with Exclusion
  Bounds from LEP and the Tevatron}.
\newblock {\em Comput. Phys. Commun.}, 181:138--167, 2010, 0811.4169.

\bibitem{Bechtle:2011sb}
Philip Bechtle, Oliver Brein, Sven Heinemeyer, Georg Weiglein, and Karina~E.
  Williams.
\newblock {HiggsBounds 2.0.0: Confronting Neutral and Charged Higgs Sector
  Predictions with Exclusion Bounds from LEP and the Tevatron}.
\newblock {\em Comput. Phys. Commun.}, 182:2605--2631, 2011, 1102.1898.

\bibitem{Bechtle:2013wla}
Philip Bechtle, Oliver Brein, Sven Heinemeyer, Oscar Stål, Tim Stefaniak,
  Georg Weiglein, and Karina~E. Williams.
\newblock {$\mathsf{HiggsBounds}-4$: Improved Tests of Extended Higgs Sectors
  against Exclusion Bounds from LEP, the Tevatron and the LHC}.
\newblock {\em Eur. Phys. J.}, C74(3):2693, 2014, 1311.0055.

\bibitem{Bechtle:2015pma}
Philip Bechtle, Sven Heinemeyer, Oscar Stal, Tim Stefaniak, and Georg Weiglein.
\newblock {Applying Exclusion Likelihoods from LHC Searches to Extended Higgs
  Sectors}.
\newblock {\em Eur. Phys. J.}, C75(9):421, 2015, 1507.06706.

\bibitem{Bechtle:2013xfa}
Philip Bechtle, Sven Heinemeyer, Oscar Stål, Tim Stefaniak, and Georg
  Weiglein.
\newblock {$HiggsSignals$: Confronting arbitrary Higgs sectors with
  measurements at the Tevatron and the LHC}.
\newblock {\em Eur. Phys. J.}, C74(2):2711, 2014, 1305.1933.

\bibitem{Goudelis:2013uca}
A.~Goudelis, B.~Herrmann, and O.~Stal.
\newblock {Dark matter in the Inert Doublet Model after the discovery of a
  Higgs-like boson at the LHC}.
\newblock {\em JHEP}, 09:106, 2013, 1303.3010.

\bibitem{ufoidm}
https://feynrules.irmp.ucl.ac.be/wiki/ModelDatabaseMainPage.
\newblock (as checked on Dec. 18,2020).

\bibitem{Zyla:2020zbs}
P.~A. Zyla et~al.
\newblock {Review of Particle Physics}.
\newblock {\em PTEP}, 2020(8):083C01, 2020.

\bibitem{gfitter}
http://project-gfitter.web.cern.ch/project-gfitter/.

\bibitem{Aghanim:2018eyx}
N.~Aghanim et~al.
\newblock {Planck 2018 results. VI. Cosmological parameters}.
\newblock 2018, 1807.06209.

\bibitem{Sirunyan:2019twz}
Albert~M Sirunyan et~al.
\newblock {Measurements of the Higgs boson width and anomalous $HVV$ couplings
  from on-shell and off-shell production in the four-lepton final state}.
\newblock {\em Phys. Rev. D}, 99(11):112003, 2019, 1901.00174.

\bibitem{ATLAS-CONF-2020-052}
{Combination of searches for invisible Higgs boson decays with the ATLAS
  experiment}.
\newblock Technical report, CERN, Geneva, Oct 2020.
\newblock All figures including auxiliary figures are available at
  https://atlas.web.cern.ch/Atlas/GROUPS/PHYSICS/CONFNOTES/ATLAS-CONF-2020-052.

\bibitem{Lundstrom:2008ai}
Erik Lundstrom, Michael Gustafsson, and Joakim Edsjo.
\newblock {The Inert Doublet Model and LEP II Limits}.
\newblock {\em Phys. Rev.}, D79:035013, 2009, 0810.3924.

\bibitem{Akerib:2013tjd}
D.~S. Akerib et~al.
\newblock {First results from the LUX dark matter experiment at the Sanford
  Underground Research Facility}.
\newblock {\em Phys. Rev. Lett.}, 112:091303, 2014, 1310.8214.

\bibitem{Kalinowski:2018kdn}
Jan Kalinowski, Wojciech Kotlarski, Tania Robens, Dorota Sokolowska, and
  Aleksander~Filip Zarnecki.
\newblock {Exploring Inert Scalars at CLIC}.
\newblock {\em JHEP}, 07:053, 2019, 1811.06952.

\bibitem{deBlas:2018mhx}
J.~de~Blas et~al.
\newblock {The CLIC Potential for New Physics}.
\newblock 3/2018, 12 2018, 1812.02093.

\bibitem{Zarnecki:2019poj}
Aleksander~Filip Zarnecki, Jan Kalinowski, Jan Klamka, Pawel Sopicki, Wojciech
  Kotlarski, Tania Robens, and Dorota Sokolowska.
\newblock {Inert Doublet Model Signatures at Future e+e- Colliders}.
\newblock {\em PoS}, ALPS2019:010, 2020, 1908.04659.

\bibitem{Zarnecki:2020swm}
Aleksander~Filip Zarnecki, Jan Kalinowski, Jan Klamka, Pawel Sopicki, Wojciech
  Kotlarski, Tania Robens, and Dorota Sokolowska.
\newblock {Searching Inert Scalars at Future e$^+$e$^-$ Colliders}.
\newblock In {\em {International Workshop on Future Linear Colliders}}, 2 2020,
  2002.11716.

\bibitem{Sokolowska:2019xhe}
Dorota Sokolowska, Jan Kalinowski, Jan Klamka, Pawel Sopicki, Aleksander~Filip
  Zarnecki, Wojciech Kotlarski, and Tania Robens.
\newblock {Inert Doublet Model signatures at future $e^+e^-$ colliders}.
\newblock {\em PoS}, EPS-HEP2019:570, 2020, 1911.06254.

\bibitem{Klamka:2022ukx}
Jan Klamka and Aleksander~Filip Zarnecki.
\newblock {Pair-production of the charged IDM scalars at high energy CLIC}.
\newblock 1 2022, 2201.07146.

\bibitem{Moretti:2001zz}
Mauro Moretti, Thorsten Ohl, and Jurgen Reuter.
\newblock {O'Mega: An Optimizing matrix element generator}.
\newblock pages 1981--2009, 2001, hep-ph/0102195.

\bibitem{Kilian:2007gr}
Wolfgang Kilian, Thorsten Ohl, and Jurgen Reuter.
\newblock {WHIZARD: Simulating Multi-Particle Processes at LHC and ILC}.
\newblock {\em Eur. Phys. J.}, C71:1742, 2011, 0708.4233.

\bibitem{Staub:2015kfa}
Florian Staub.
\newblock {Exploring new models in all detail with SARAH}.
\newblock {\em Adv. High Energy Phys.}, 2015:840780, 2015, 1503.04200.

\bibitem{Porod:2003um}
Werner Porod.
\newblock {SPheno, a program for calculating supersymmetric spectra, SUSY
  particle decays and SUSY particle production at e+ e- colliders}.
\newblock {\em Comput. Phys. Commun.}, 153:275--315, 2003, hep-ph/0301101.

\bibitem{Linssen:2012hp}
Lucie Linssen, Akiya Miyamoto, Marcel Stanitzki, and Harry Weerts.
\newblock {Physics and Detectors at CLIC: CLIC Conceptual Design Report}.
\newblock 2012, 1202.5940.

\bibitem{Hocker:2007ht}
Andreas Hocker et~al.
\newblock {TMVA - Toolkit for Multivariate Data Analysis}.
\newblock 2007, physics/0703039.

\bibitem{Delahaye:2019omf}
Jean~Pierre Delahaye, Marcella Diemoz, Ken Long, Bruno Mansoulié, Nadia
  Pastrone, Lenny Rivkin, Daniel Schulte, Alexander Skrinsky, and Andrea
  Wulzer.
\newblock {Muon Colliders}.
\newblock 2019, 1901.06150.

\bibitem{Robens:2021zvr}
Tania Robens, Jan Kalinowski, Aleksander~Filip Zarnecki, and Andreas
  Papaefstathiou.
\newblock {Extended scalar sectors at future colliders}.
\newblock In {\em {27th Cracow Epiphany Conference on Future of particle
  physics }}, 3 2021, 2104.00046.

\bibitem{Ipek:2014gua}
Seyda Ipek, David McKeen, and Ann~E. Nelson.
\newblock {A Renormalizable Model for the Galactic Center Gamma Ray Excess from
  Dark Matter Annihilation}.
\newblock {\em Phys. Rev.}, D90(5):055021, 2014, 1404.3716.

\bibitem{No:2015xqa}
Jose~Miguel No.
\newblock {Looking through the pseudoscalar portal into dark matter: Novel
  mono-Higgs and mono-Z signatures at the LHC}.
\newblock {\em Phys. Rev.}, D93(3):031701, 2016, 1509.01110.

\bibitem{Goncalves:2016iyg}
Dorival Goncalves, Pedro A.~N. Machado, and Jose~Miguel No.
\newblock {Simplified Models for Dark Matter Face their Consistent
  Completions}.
\newblock {\em Phys. Rev.}, D95(5):055027, 2017, 1611.04593.

\bibitem{Bauer:2017ota}
Martin Bauer, Ulrich Haisch, and Felix Kahlhoefer.
\newblock {Simplified dark matter models with two Higgs doublets: I.
  Pseudoscalar mediators}.
\newblock {\em JHEP}, 05:138, 2017, 1701.07427.

\bibitem{Tunney:2017yfp}
Patrick Tunney, Jose~Miguel No, and Malcolm Fairbairn.
\newblock {Probing the pseudoscalar portal to dark matter via $\bar
  bbZ(\to\ell\ell)+ \not{E}_T$ : From the LHC to the Galactic Center excess}.
\newblock {\em Phys. Rev.}, D96(9):095020, 2017, 1705.09670.

\bibitem{LHCDarkMatterWorkingGroup:2018ufk}
Tomohiro Abe et~al.
\newblock {LHC Dark Matter Working Group: Next-generation spin-0 dark matter
  models}.
\newblock {\em Phys. Dark Univ.}, 27:100351, 2020, 1810.09420.

\bibitem{Robens:2021lov}
Tania Robens.
\newblock {The THDMa Revisited}.
\newblock {\em Symmetry}, 13(12):2341, 2021, 2106.02962.

\end{thebibliography}


\begin{thebibliography}{10}



\end{thebibliography}

\end{document}